\documentclass[aps,prl,reprint,superscriptaddress]{revtex4-1}
\usepackage{blindtext}
\usepackage{graphicx}
\usepackage{amsmath,amssymb}
\usepackage{multirow}
\usepackage{url}
\usepackage[colorlinks,
            linkcolor=blue,
            anchorcolor=blue,
            citecolor=blue,
            ]{hyperref}
\usepackage{setspace}

\begin{document}

\title{Classification and prediction of wave chaotic systems with machine learning techniques}

\author{Shukai Ma}
\affiliation{Department of Physics, University of Maryland, College Park, Maryland 20742-4111, USA}
\author{Bo Xiao}
\affiliation{Department of Electrical and Computer Engineering, University of Maryland, College Park, Maryland 20742-3285, USA}
\author{Ron Hong}
\affiliation{U.S. Naval Research Laboratory, Washington, DC 20375, USA}
\author{Bisrat D. Addissie}
\affiliation{U.S. Naval Research Laboratory, Washington, DC 20375, USA}
\author{Zachary B. Drikas}
\affiliation{U.S. Naval Research Laboratory, Washington, DC 20375, USA}
\author{Thomas M. Antonsen}
\affiliation{Department of Physics, University of Maryland, College Park, Maryland 20742-4111, USA}
\affiliation{Department of Electrical and Computer Engineering, University of Maryland, College Park, Maryland 20742-3285, USA}
\author{Edward Ott}
\affiliation{Department of Physics, University of Maryland, College Park, Maryland 20742-4111, USA}
\affiliation{Department of Electrical and Computer Engineering, University of Maryland, College Park, Maryland 20742-3285, USA}
\author{Steven M. Anlage}
\affiliation{Department of Physics, University of Maryland, College Park, Maryland 20742-4111, USA}
\affiliation{Department of Electrical and Computer Engineering, University of Maryland, College Park, Maryland 20742-3285, USA}

\begin{abstract}
The wave properties of complex scattering systems that are large compared to the wavelength, and show chaos in the classical limit, are extremely sensitive to system details. A solution to the wave equation for a specific configuration can change substantially under small perturbations. Due to this extreme sensitivity, it is difficult to discern basic information about a complex system simply from scattering data as a function of energy or frequency, at least by eye. In this work, we employ supervised machine learning algorithms to reveal and classify hidden information about the complex scattering system presented in the data. As an example we are able to distinguish the total number of connected cavities in a linear chain of weakly coupled lossy enclosures from measured reflection data. A predictive machine learning algorithm for the future states of a perturbed complex scattering system is also trained with a recurrent neural network.  Given a finite training data series, the reflection/transmission properties can be forecast by the proposed algorithm.
\end{abstract}

\maketitle

\section {I. Introduction}

The scattering of short-wavelength waves inside large enclosures is commonly encountered in various fields of physics and engineering, including quantum dots \cite{Alhassid2001}, optical cavities \cite{Notomi2008}, and microwave enclosures \cite{Dietz2006}. In wave chaotic systems the trajectory of the short-wavelength rays shows chaotic dynamics \cite{ott_2002,Dietz2016}. Minute perturbations of the system, such as the enclosure boundary condition changes, the shift of the driving frequency, and the environmental conditions, can drastically change the wave properties \cite{Dupre2015, Kaina2015, DelHougne2018}. On the other hand, it is very difficult to extract structural or configurational information from raw system measurements since the effect of structural changes on the wave trajectories is effectively hidden by the ray-chaotic nature of the system.
Deterministic numerical methods are not suitable for studying wave chaotic systems, since the small wavelength feature calls for extremely dense meshing \cite{Parmantier2004}, and a small change in the system configuration may require a completely new solution each time. 

Statistical approaches have been proposed to understand wave chaotic systems, such as the Dynamic Energy Analysis method \cite{Tanner2009, Bajars2017}, the Power Balance Method \cite{Hill1994, Hill1998,Bajars2017, Junqua2005}, and the Random Coupling Model (RCM) \cite{Wigner1955,Yeh,Zheng2006,Zheng2006a,Hart2009,Gradoni2014,Xiao2016}. The RCM accounts for the system-specific features such as the coupling between ports and the enclosure, and the universal underlying fluctuations described by random matrix theory, which are conjectured to describe the chaotic behaviour of the wave chaotic system \cite{McDonald1979,Casati1980,Berry1981,Bohigas1984}. RCM can generate both mean-field and statistical predictions utilizing a minimum of information, namely the system coupling details and the enclosure loss parameter $\alpha$ \cite{Zheng2006,Zheng2006a,Gradoni2012,Li2015, Xiao2018}. 

Machine learning (ML) techniques have enjoyed intense research interest in recent years \cite{Mehta2018,DasSarma2019}. A ML algorithm treats any given task as a mathematical problem and does not utilize knowledge of the specific physics underlying the data. Benefiting from this `model-free' nature, the ML algorithms find broad application in various sub-fields of physics, such as the identification of phase transitions in condensed matter studies \cite{Carrasquilla2017, Venderley2018}, the classification of multi-qubit states of trapped-ion experiments \cite{Seif2018}, the auto-tuning of gate voltages in quantum dots system \cite{VanDiepen2018, Kalantre2019}, and the future state predictions of spatio-temporal chaotic systems \cite{Lu2017,Pathak2018,Neofotistos2019}. Although successfully applied in various studies, one crucial drawback of the ML techniques is the trade-off between a successfully trained program and the amount of data required during its training phase. However, the problem of training data acquisition does not pose a great challenge to the analysis of wave chaotic systems. In order to compare data to theoretical predictions based on the statistical methods mentioned in the previous paragraph, a large statistical ensemble of measured data is required. This feature of wave chaotic system analysis is suitable for building a successfully trained machine learning algorithm.

In this manuscript, we utilize machine learning techniques to deepen our understanding of wave chaotic systems. In particular we would like to classify the detailed underlying ``hidden'' properties of a wave chaotic system, and to predict its future states when subjected to systematic perturbations. The paper is organized as follows: We start with an introduction to the cascaded chaotic cavity experiment in section II. The machine learning algorithms are discussed in section III, and the results of the ML program outputs are shown and discussed in section IV. We conclude the work and discuss future directions in section V.

\section {II. experimental set-up of a Wave chaos system}

\begin{figure}
\centering
\includegraphics[width=0.5\textwidth]{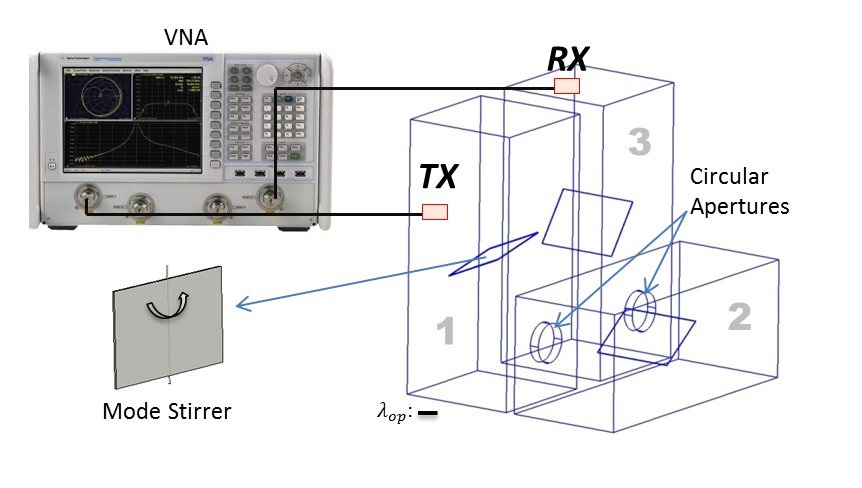}
\caption{\label{fig:exp} Schematic of the experimental set-up. We measure the S-parameter of a 3-cavity cascade system with a VNA. The cavities are connected through circular apertures. Rotatable mode stirrers are employed in each cavity to generate different system configurations. The scale of the operating wavelength $\lambda_{op}$ is shown as the bar. }
\end{figure}

We study the transmission and reflection of electro-magnetic (EM) waves in wave chaotic enclosures. Short-wavelength EM waves from 3.95-5.85 GHz are injected into cavities of dimension $0.762\times0.762\times1.778 m^3$ through WR178 single-mode waveguides. The cavities are electrically large ($\sim 10^4$ modes at the operating frequency range) in order to simulate real-life examples of wave chaotic settings such as rooms in buildings and cabins in a ship. A series of individual cavities can be connected into a linear cascade chain through apertures as shown in Fig. \ref{fig:exp}. The total number of connected cavities is varied from 1 to 3. Independent mode stirrers are employed inside each cavity to create a large ensemble of statistically distinct realizations of the system \cite{Frazier2013,Frazier2013a,Hemmady,Drikas2014}. Single-mode ports are mounted on the first and last cavity in the cascade, shown as T(R)X in Fig. \ref{fig:exp}. 
The $2\times2$ Scattering(S)-parameters of the entire cavity cascade system are measured with a Vector Network Analyzer (VNA), and the $2\times2$ impedance(Z)-matrix calculated. The S and Z parameters are connected through the bilinear transformation, $\underline{\underline{S}}=\underline{\underline{Z_0}}^{0.5}(\underline{\underline{Z}}+\underline{\underline{Z_0}})^{-1}(\underline{\underline{Z}}-\underline{\underline{Z_0}})^{-1}\underline{\underline{Z_0}}^{0.5}$, where $\underline{\underline{Z_0}}$ is a diagonal matrix whose elements are the characteristic impedance of the waveguide channels leading to the ports. All mode stirrers are rotated simultaneously to ensure a low correlation between each transmission measurement, and the S-parameter measurements are carried out when the mode stirrers are stationary. A total number of 200 distinct realizations of the cavity cascade are created.
The degree of loss in each cavity is altered by placing RF absorber cones in each cavity. The single cavity ``lossyness'' is described by the RCM loss parameter $\alpha$ which is defined as the ratio of the 3-dB bandwidth of a resonance mode to the mean spacing between the modes \cite{Gradoni2014, Xiao2018} and is basically the number of overlapping modes at a given frequency. The loss parameter can have values ranging from 0 (no loss) to infinity (extremely lossy).

\begin{figure*}
\centering
\includegraphics[width=0.91\textwidth]{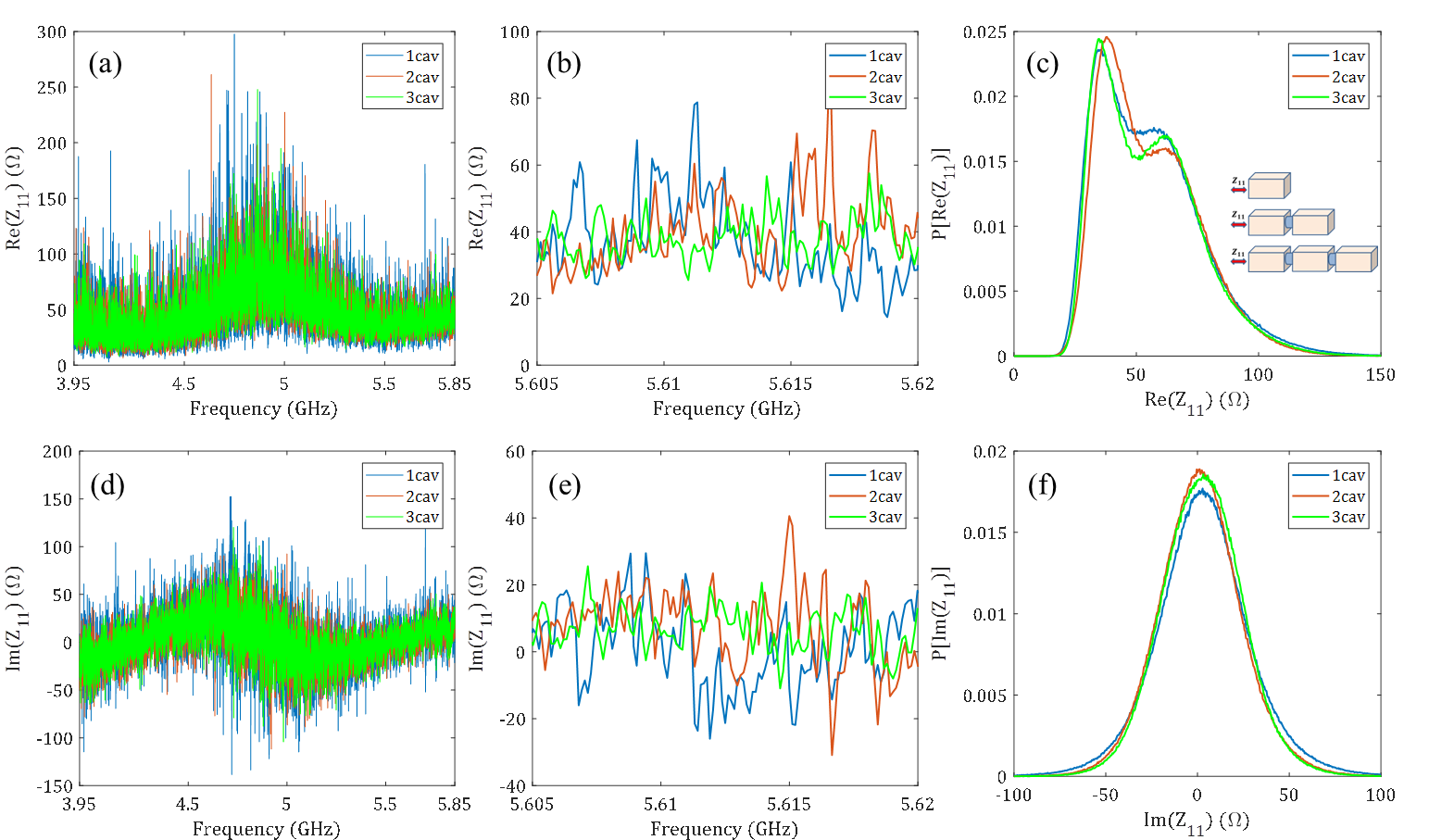}
\caption{\label{fig:z11data} Multi-cavity diagonal impedances $Z_{11}$ from 2-port measurements on either 1-cavity (blue), 2-cavity (red) or 3-cavity (green) cascade. 
The inset in (c) is the schematic diagram of the Z-parameter measurement of 1-, 2- and 3-cavity cascade structures. (a) and (d): the real and imaginary parts of $Z_{11}$ from 3.95-5.85GHz for a single realization of the system, whose detailed views are shown in (b) and (e), and the corresponding statistical analysis over a large ensemble of $Re(Z_{11})$ and $Im(Z_{11})$ are plotted in (c) and (f). The single cavity loss parameter is $\alpha=9.7$.}
\end{figure*}

The real and imaginary parts of diagonal impedance $Z_{11}$ measurements for a single realization is shown in Fig. \ref{fig:z11data}. The curves correspond to systems with a different number of connected cavities, varying from 1 to 3. The cavities are nominally identical, in that all cavities share the same physical dimension and a uniform single cavity loss parameter value: $\alpha=9.7$. It has been demonstrated both theoretically and experimentally that the statistical properties of the diagonal impedance $Z_{11}$ of a high-loss ($\alpha>>1$) cavity cascade system remain virtually unchanged regardless of the total number of connected cavities in the chain \cite{Gradoni2012}. The real and imaginary parts of measured $Z_{11}$ values of multi-cavity systems are shown in Fig. \ref{fig:z11data} (a) and (d). Direct observation of the frequency dependent $Z_{11}$ results cannot provide useful information to classify the number of connected cavities, since the curves are essentially on top of each other and effectively indistinguishable vs. frequency. The detailed $Re(Z_{11})$ and $Im(Z_{11})$ from 5.605-5.62GHz are shown in Fig. \ref{fig:z11data} (b) and (e). Even with the detailed view, it is still hard to differentiate the $Z_{11}$ curves from either the mode density, the level of fluctuation, or the averaged $Z_{11}$ value levels. Statistical analysis of the $Re(Z_{11})$ and $Im(Z_{11})$ measurements are shown in \ref{fig:z11data} (c) and (f). The $Z_{11}$ PDFs of 1-, 2- and 3-cavity cascade are very similar to each other and difficult to systematically distinguish. The Fourier transforms of the multi-cavity $Z_{11}$ data show similar time-domain responses.
We first wish to see if an ML algorithm can be trained to correctly distinguish the number of cavities in the cascade simply from raw data such as that shown in Fig. \ref{fig:z11data}. The second objective is to see if an ML algorithm can predict the evolution of the S- (or Z-) parameters as the cavity cascade is systematically perturbed.

\section {III. ML model: neural network}

\begin{figure}
\centering
\includegraphics[width=0.5\textwidth]{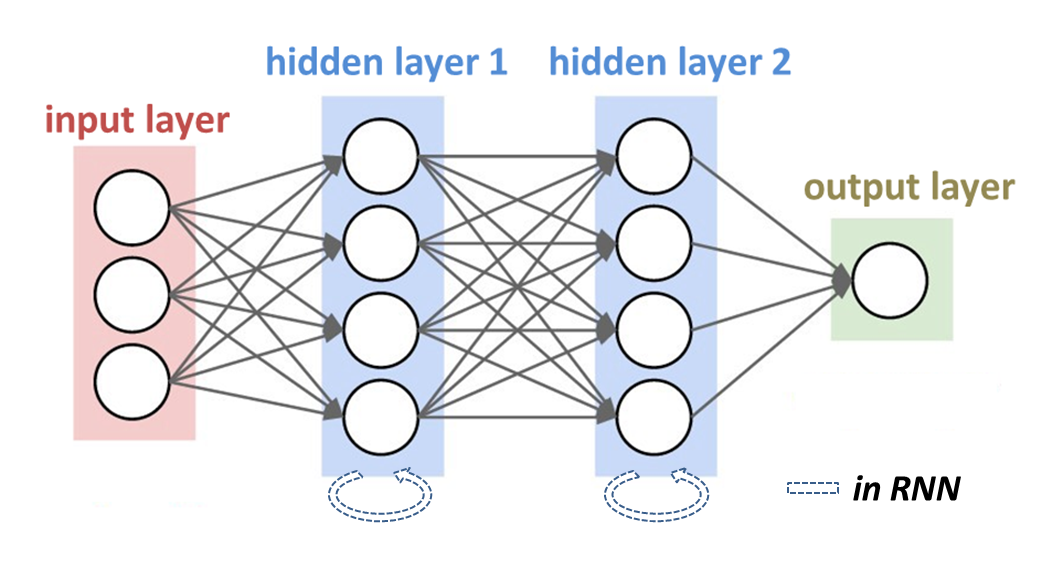}
\caption{\label{fig:nn} Generalized (recurrent) neural network architecture with two hidden layers. The NN consists of the one input layer, hidden layers and one output layer. The RNN is built upon the NN structure with the addition of ``memory effects'' at each hidden layer (shown as the dashed arrows).}
\end{figure}


The neural network (NN) and its modifications, inspired by the electric signal propagation mechanism of brain neuron cells, are widely applied in various fields, such as speech identification, pattern recognition and picture captioning \cite{Lecun2015}. A neural network is one of the supervised learning algorithms whose goal is to utilize the given training set information and establish a generic method to assign labels to the testing sets. As a typical example of classifying cat and dog pictures, a supervised learning algorithm refines its classification rule using thousands of images of correctly labeled cat and dog images. More specifically, the algorithm identifies the label of a given picture through the images pixel values. A typical structure of a NN is shown in Fig. \ref{fig:nn}. We will next briefly introduce the operating principle of the NN with the example of an $m-$class picture classification task ($m=2$ for the cat and dog classification example). 

In each iteration, the information of one picture from the training set is prepared into a column vector and used as the input of the NN algorithm. For example, a picture of $L\times L$ pixels is reshaped into a $(L^2,1)$ vector $x$ by simply concatenating the $L\times1$ lines $L$ times. 
The input vector is further modified as $\Tilde{x}=\frac{x-<x>}{\sigma_x}$ in order to improve the performance of the algorithm, where $<x>$ and $\sigma_x$ are the mean and standard derivation of $x$. 
After the input preparation, the $i$'th input $\Tilde{x}^{(i)}$ is passed on to the first (represented in the subscripts) hidden layer $h^{(i)}_1$ (with $n_1$ units) through $h^{(i)}_1=\sigma(W^{(i)}_{1}\cdot \Tilde{x}^{(i)} + b^{(i)}_1)$, where $W^{(i)}_{1}$ and $b^{(i)}_1$ are a $(n_1, L^2)$ coupling matrix and a $(n_1,1)$ bias vector, respectively. 
The non-linear function $\sigma(\cdot)$ adds to the complexity of the NN and further expands the network's functionality. 
The first hidden layer vector $h^{(i)}_1$ is transferred to the second hidden layer through $h^{(i)}_2=\sigma(W^{(i)}_2\cdot h^{(i)}_1 + b^{(i)}_2)$, where $W^{(i)}_2$ and $b^{(i)}_2$ are the coupling matrix and the bias of the second layer, and the same nonlinear function $\sigma(\cdot)$ is used. 
The algorithm labels the input picture using $y^{(i)}_{NN}=f[\sigma(W^{(i)}_{n+1}\cdot h^{(i)}_n + b^{(i)}_{n+1})]$. Similarly, $W^{(i)}_{n+1}$ and $b^{(i)}_{n+1}$ are the coupling matrix and bias at the output coupling layer, and $f$ is a normalization function which transforms the value of $\sigma(W^{(i)}_{n+1}\cdot h^{(i)}_n + b^{(i)}_{n+1})$ into a `one-hot' label vector. 
The `one-hot' label is a one-column vector with $m$ elements representing the $m$ possible output classes. For a given training data input, the normalization function $f$ sets only one vector element of $y_{NN}^{(i)}$ to 1, which represents the label of the class for that particular data set, and the other elements to zero. In the meantime, a cost function $f_{cost}=||y^{(i)}_{NN}-y^{(i)}||$ is calculated to describe the distance between the algorithm label $y^{(i)}_{NN}$ and the known label $y^{(i)}$ for the given input $x^{(i)}$. The goal of the training process is to minimize the value of $f_{cost}$ through the so-called back propagation process, where the algorithm refines the values of all coupling matrices $W$ and biases $b$, utilizing the derivative values $\frac{df_{cost}}{dW}$ and $\frac{df_{cost}}{db}$. The back propagation process is repeated for all pictures in the training set. The values of all network parameters are fixed after the training. In the testing phase, unseen pictures are fed into the trained network with their machine predicted labels calculated, while the back propagation process is not used. If the predicted labels agree well with the correct labels, the classification NN algorithm is successfully trained.

\begin{table*}
\begin{ruledtabular}
\begin{tabular*}{\textwidth}{@{\extracolsep{\fill}}lcc}
  & NN & RNN\\
  \hline
Forward data flow & $x^{(i)} \rightarrow h^{(i)} \rightarrow y^{(i)}$ & $x^{(i)} \rightarrow h^{(i)} (\circlearrowright h^{(i-1)}) \rightarrow y^{(i)} $\\
Hidden layer & $h^{(i)}=\sigma (W^{(i)}_{x\rightarrow h}\cdot x^{(i)} + b^{(i)})$ & $h^{(i)}=\sigma (W^{(i)}_{x\rightarrow h}\cdot x^{(i)} + W^{(i)}_{h\rightarrow h}\cdot h^{(i-1)} + b^{(i)})$ \\
Output & $y^{(i)}=W^{(i)}_{h\rightarrow y} \cdot h^{(i)}$ & $y^{(i)}=W^{(i)}_{h\rightarrow y} \cdot h^{(i)}$ \\
\end{tabular*}
\end{ruledtabular}
\caption{\label{tab:catalog} The 1-hidden-layer NN and RNN comparison. $x, h$ and $y$ are to the input, hidden layer and output vectors, respectively. $\sigma$ is the non-linear function. The superscript $i$ refers to the index of input series. $b$ is the bias vector and $W_{k\rightarrow j}$ refers to the coupling matrix from vector $k$ to vector $j$. }
\end{table*}

The recurrent neural network (RNN) structure is based on the neural network with the addition of a memory effect. 
As shown in Fig. \ref{fig:nn}, extra loops are employed in every hidden layer. We use table \ref{tab:catalog} to illustrate the difference between a 1-hidden-layer NN and a 1-hidden-layer RNN. The RNN hidden layer $h^{(i)}$ has an additional influence from its previous state $h^{(i-1)}$ through the $h \rightarrow h$ coupling matrix $W^{(i)}_{h\rightarrow h}$. This feature grants a RNN the ability to store information from previous iterations, and the potential to predict future states. Differing from the NN algorithm, which is insensitive to the specific order of the input information, a RNN algorithm requires that the inputs are fed into the algorithm following a strict time-ordering or systematic evolution. 

\section {IV. Wave Chaos Case studies}

\subsection{Cascaded multi-cavity system classification}

\begin{figure}
\centering
\includegraphics[width=0.51\textwidth]{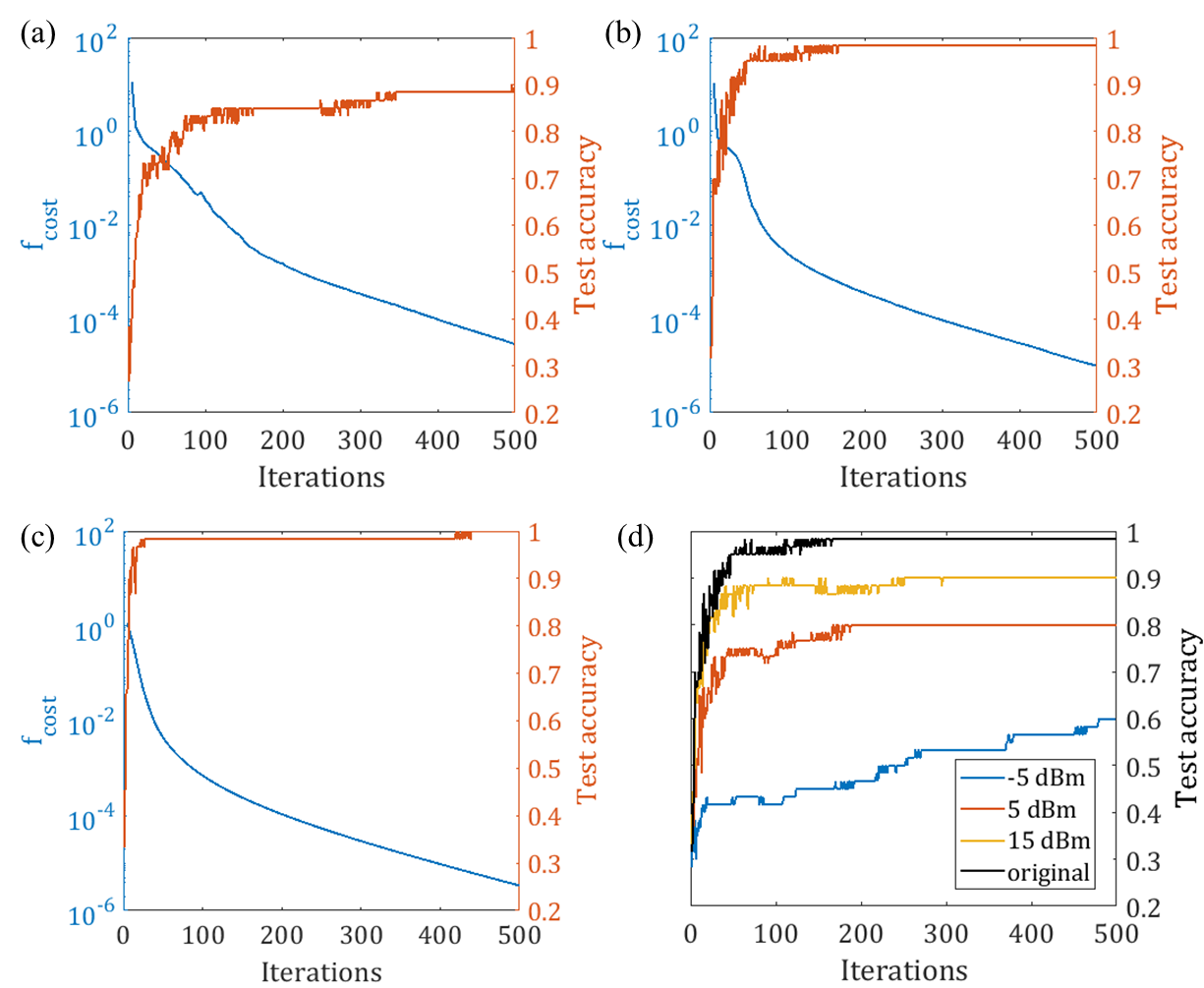}
\caption{\label{fig:train} The NN classification algorithm performance for 1-, 2-, and 3-cavity $Re(Z_{11})$ data. (a-c): The cost function $f_{cost}$ (blue, left logarithmic axis) and test accuracy (red, right linear axis) results are evaluated to 500 iterations of the algorithm. The dimension of the input vector is 2000, 5000 and 16001 in (a), (b) and (c), respectively. (d): The test accuracy results with and without additional Gaussian noise. The dimension of the input vector is 5000. The signal to noise ratio values are -5, 5, 15 dB and the original data.}
\end{figure}

We utilize the neural network algorithm to classify different wave chaotic systems based on raw scattering data. The goal is to train the machine to classify the total number of cavities in the linear cascade chain by using the raw impedance (Z) matrix measurements. As discussed in section I, the diagonal element $Z_{11}$ of 1-, 2- and 3-cavity systems is very difficult to distinguish by eye, and also shares nearly identical statistical properties as shown in Fig. \ref{fig:z11data}. In the 1-, 2- and 3-cavity cascade measurements, we rotate the mode stirrer to 200 unique and distinct angles in order to generate a large ensemble of measured data. The impedance measurement sweep is from 3.95-5.85 GHz with 16001 frequency points. One-hot vectors of size $(3,1)$ are created as the correct labels for all configurations. $80\%$ of the total measurements ($3\times 200=600$ realizations from 1-, 2- and 3-cavity cascade cases) are used to train the algorithm, and $20\%$ of the data is reserved as a testing data set. The NN employs 4 hidden layers which have 25, 26, 33 and 18 units, respectively. The specific choice of the total number of layers and units per layer can be varied. We use the Tensorflow \cite{tensorflow2015-whitepaper} package in the back propagation process. The training and testing results are shown in Fig. \ref{fig:train}. To speed up the algorithm, we select subsets of size 2000 (Fig. \ref{fig:train} (a)) and 5000 (Fig. \ref{fig:train} (b)) uniformly chosen frequency points out of the total 16001 $Re(Z_{11})$ measurements as input vectors (the $Im(Z_{11})$ input algorithm also works just as well but the results are not shown). In both cases, the cost function decreases sharply and the testing set accuracy reaches above $95\%$ within 500 cost-function minimization iterations. We observe that training the algorithm with 5000 data points in the input vectors (Fig. \ref{fig:train} (b)) reaches a higher testing accuracy (98.3\%) as compared to the fewer data point case (95\%). The testing accuracy reaches to 100\% (Fig. \ref{fig:train} (c)) when all measured data points are fed into the NN algorithm.
One can understand the improvement of test performance from the fact that more information are delivered to the algorithm with the increase of input vector size. The algorithm successfully identifies the total number of connected cavities in the cascade utilizing only the knowledge of the system diagonal impedance as a function of frequency. The result indicates that the algorithm is able to detect and utilize ``hidden order'' embedded in the measured data, completing a task that is hard to achieve with either visual inspection or by more sophisticated analysis and statistical means \cite{Venderley2018}.

We further explore the limitations and robustness of the developed multi-cavity classification algorithm. We first test the algorithm's ability to identify unseen measured data. An algorithm is trained to successfully classify the $Re(Z_{11})$ measured data from the 1- and 3-cavity cascade systems. The NN output has three classes: the 1- and 3-cavity systems whose data are seen by the algorithm in the training phase, and an untrained 2-cavity system class. In the testing phase, we mix in the $Re(Z_{11})$ data from 2-cavity cascade measurements. We find that the algorithm classifies the 2-cavity data into the all three categories with equal probabilities. This result indicates that the classification algorithm applies strictly to the data drawn from the training ensemble. We next test the algorithm's tolerance level to noise. The algorithm is successfully trained to classify the $Re(Z_{11})$ data from 1-, 2-, and 3-cavity cascade measurements with minimal experimental noise. In the testing phase, Gaussian noise are added to the $Re(Z_{11})$ measurements to achieve signal to noise ratios (SNR) of -5, 5 and 15 dB. The power level of the noise $P_n$ is calculated from $SNR=10\log_{10}(P_s/P_n)$ based on the signal power level $P_s$.
As shown in Fig. \ref{fig:train} (d), the final test accuracy changed from 0.9 to 0.66 (saturating after the 1000 iterations) with the decrease of SNR from +15 to -5 dB. The result shows that the classification algorithm retains its ability to distinguish the cavity number when the noise level is not too high. 

The use of the frequency and realization-averaged $Z_{11}$ values was investigated as a possible alternative to the machine learning algorithm. The 1-, 2- and 3-cavity cascade $\left< Re(Z_{11}) \right>$ are nearly identical (varying by 0.7\% of the average value). The ML algorithm fully retains the classification performance for the multi-cavity $Re(Z_{11})$ data even after it has been normalized by the average values. This demonstrates that the algorithm utilizes details that are not easily summarized when making a high resolution distinction between the three scattering scenarios based on statistics. With statistical methods, repetitive independent measurements of the multi-cavity system can produce stable mean values of $\left< Re(Z_{11}) \right>$ to further distinguish the different cases. To obtain a good estimation of $\left< Re(Z_{11}) \right>$ and decrease the uncertainty, this approach involves a large number of measurements of each different system in the data-gathering phase. In the prediction phase, given an unknown system,  a large amount of additional data is required to acquire a good measurement of $\left< Re(Z_{11}) \right>$ to perform classification. By contrast, a trained ML algorithm can distinguish an unknown system with only one round of measurement. Compared to the statistical method, the ML algorithm requires considerably less data to produce a confident classification of the system.

\subsection{Chaotic cavity S-parameter series prediction with a RNN}

Inspired by the recent progress of predicting the future evolution of classically chaotic systems with machine learning techniques \cite{Lu2017,Pathak2018,Neofotistos2019}, we utilize the RNN algorithm to predict the evolution of a wave chaotic system under systematic perturbation. We record the $2\times 2$ S-parameter matrix of a single wave chaotic enclosure under a sequential and systematic perturbation from a rotating mode stirrer. The dimension of the single cavity is $0.038\times 0.038 \times 0.089m^3$. EM waves from 75-110GHz are fed into the cavity ($\sim 10^4$ modes at the operating frequency range) through single-mode WR10 waveguides from Virginia Diodes VNA Extenders and the S-parameters are measured by a VNA.
The mode stirrer is rotated to 1000 unique angles with uniform step angle size. A full rotation of the mode stirrer takes $\sim 120$ steps. There exists some correlations between the measurements.
The task is to use a portion of the measured $S_{11}$ and $S_{21}$ data to train the RNN algorithm, and then generate predictions of $S_{21}$ by supplying additional measured $S_{11}$ data obtained later in the sequence. It is difficult to predict the future state of a wave chaotic system based solely on its history information, since minute changes of the system's boundary condition can drastically affect the EM wave properties \cite{Hemmady2012}. 

\begin{figure}
\centering
\includegraphics[width=0.5\textwidth]{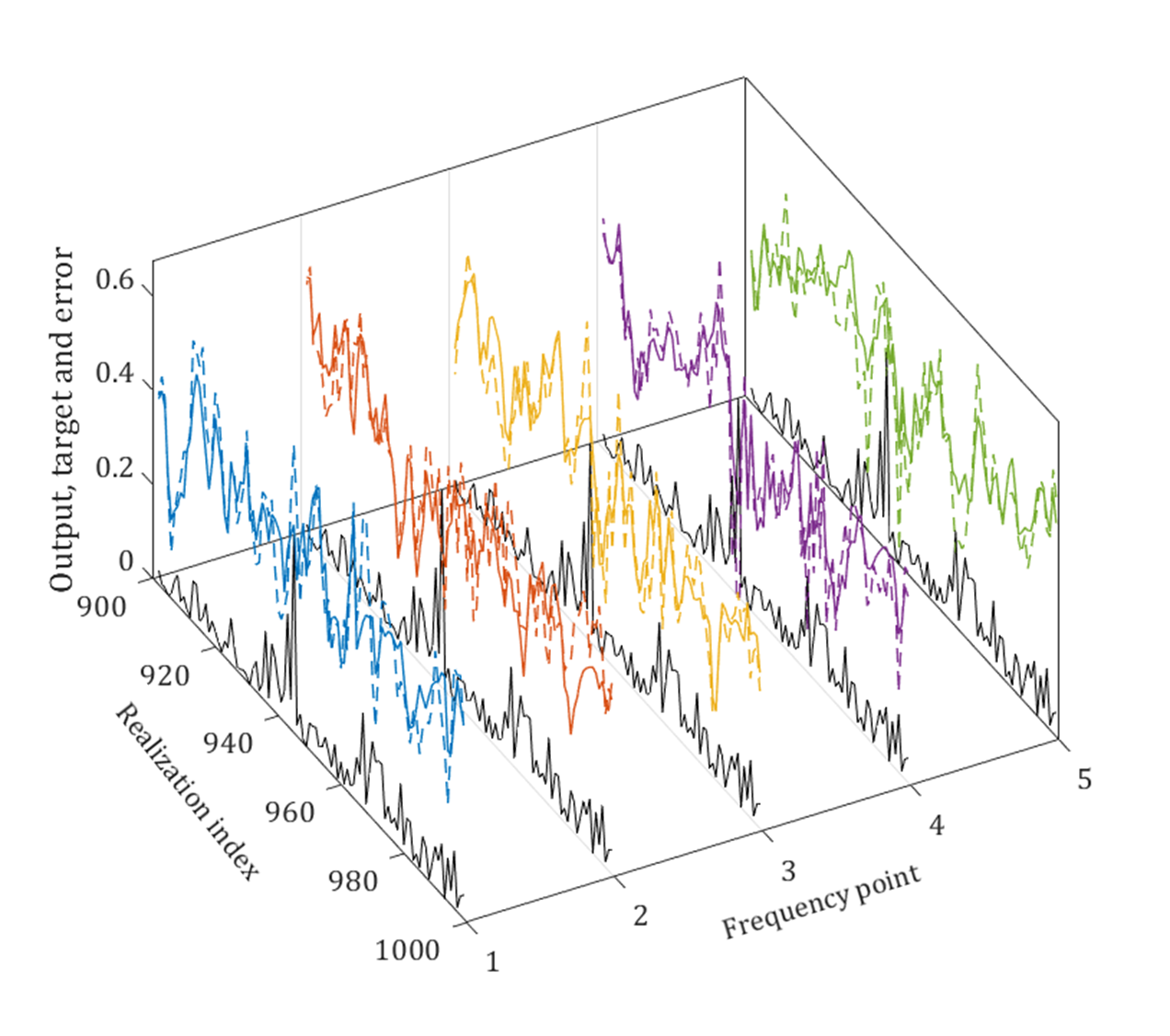}
\caption{\label{fig:predict} The $|S_{21}|$ prediction of a wave chaotic electromagnetic system future states using measured $|S_{11}|$. The measured $|S_{21}|$ data (colored solid lines) and algorithm predictions (colored dashed lines) from 5 frequency points are compared for cavity realizations 900 to 1000. The black lines show the absolute magnitude of the difference between the measured data and the algorithm prediction.}
\end{figure}

The RNN algorithm is implemented with the layer recurrent neural network function from the Matlab Deep Learning Toolbox \cite{Matlab}. Similar to the neural network training, we first prepare the input vectors $\Tilde{x}=\frac{x-<x>}{\sigma_x}$ to improve the performance of the algorithm. In contrast with the input labeling process in the NN, one must prepare the input and output vectors in the correct evolutionary order. We next define the desired network parameters, such as the sizes of the hidden layer units and the method of back-propagation optimization. In the chaotic cavity S-parameter prediction task, the input vectors are assembled from measured $|S_{11}|$ at 50 frequency points with 175MHz spacing from 75-110 GHz. The output vectors are from measured $|S_{21}|$ at 5 frequency points with 2.5GHz spacing ($\sim1000$ modes exist in this frequency interval). It is not shown here, but the algorithm also works for the real/imaginary parts of the S-parameters as input and output. The adopted RNN structure has 38 units in 1 hidden layer. The Levenberg-Marquardt method is applied in the back-propagation process to optimize the weight and bias values \cite{Levenberg1944,Marquardt1963}. The testing results from all frequency points are shown in Fig. \ref{fig:predict}. The first 900 realization of the measured data are used as the training set, and the testing begins with the 901'st realization. Only measured reflection information ($|S_{11}|$) is fed to the algorithm, serving as the observer of the system \cite{Lu2017,Neofotistos2019}, and we record the machine predicted transmission information ($|S_{21}|$). As shown in Fig. \ref{fig:predict}, we observe good agreement between the algorithm-generated and measured transmission $|S_{21}|$ at least for the first 40 or so realizations for all 5 frequency points. The prediction degrades beyond this point.

\subsection{Discussion}

The ML techniques are shown to successfully classify different wave chaotic systems and to give predictions of future system states. Benefiting from its model-free property, ML can be versatile for various tasks. However, the training process of ML methods requires large amounts of measured or simulated data, and the tuning of algorithm hyperparameters, including the number of layers, the number of units per layer, etc. These choices must be guided by experience with the algorithms rather than knowledge of the physical problem. The NN based wave chaos classification task can be further developed into a cavity loss parameter $(\alpha)$ detector, or an algorithm to distinguish different types of perturbation of a wave chaotic system, for example. The application of RNN and other techniques, such as the long short-term memory methods, the nonlinear autoregressive neural network and reservoir computing, may give predictions for the future evolution of the S-matrix when the boundary condition is subject to systematic perturbation. The algorithm could be utilized to predict the future S-matrix of an evolving system, thus allowing identification of coherent perfect absorption conditions of a wave chaotic system \cite{Li2017}, for example.

\section {V. Conclusion}

In this paper, we show that machine learning techniques are able to harness hidden features of raw wave chaotic data sets to make subtle distinctions between different scattering systems. ML techniques may also be able to make future state predictions of the measured data in a systematically-perturbed wave chaotic system. The robustness of the classification neural network algorithm is studied and shown to be loyal to the data drawn from the training ensemble. The classification algorithm also has some degree of tolerance to the addition of noise.
The advantages and weakness of ML techniques and possible future applications are also discussed.

\begin{acknowledgments}
This work was supported by ONR under Grant No. N000141512134, AFOSR COE Grant FA9550-15-1-0171.
\end{acknowledgments}


%

\end{document}